\documentclass[english,reprint,aps,prl, superscriptaddress]{revtex4-1}
\usepackage[T1]{fontenc}
\usepackage[latin9]{inputenc}
\usepackage{amsmath}
\usepackage{amssymb}
\usepackage{graphicx}
\usepackage{changes}
\usepackage{siunitx}
\usepackage{float}
\usepackage{cleveref}

\usepackage{braket}
\usepackage[raggedright]{titlesec}

\usepackage{titlesec}
\titleformat*{\section}{\normalsize\bfseries}
\titleformat*{\subsection}{\normalsize\bfseries}
\titlespacing*\section{0pt}{12pt plus 4pt minus 2pt}{0pt plus 2pt minus 2pt}
\titlespacing*\subsection{0pt}{12pt plus 4pt minus 2pt}{0pt plus 2pt minus 2pt} 

\usepackage{upgreek}

\usepackage{xcite}
\usepackage{xr}
\makeatletter
\newcommand*{\addFileDependency}[1]{
  \typeout{(#1)}
  \@addtofilelist{#1}
  \IfFileExists{#1}{}{\typeout{No file #1.}}
}
\makeatother

\newcommand*{\myexternaldocument}[1]{%
    \externaldocument{#1}%
    \addFileDependency{#1.tex}%
    \addFileDependency{#1.aux}%
}

\myexternaldocument{npj_supplementary}

\usepackage{ragged2e}
\usepackage{subcaption}   
\DeclareCaptionJustification{justified}{\justifying}
\newcommand\nref[1]{\text{Fig. }\ref{#1}}

\makeatletter

\newcommand{\toadd}[1]{#1}
\newcommand{\toremove}[1]{}

\begin{document}

\title{High Harmonic Generation in Two-Dimensional Mott Insulators}
\author{Christopher~Orthodoxou}
\author{Amelle~Za{\"i}r}
\author{George~H.~Booth}
\email{george.booth@kcl.ac.uk}
\affiliation{Department of Physics, King's College London, Strand, London, WC2R 2LS, U.K.}

\begin{abstract}
With a combination of numerical methods, including quantum Monte Carlo, exact diagonalization, and a simplified dynamical mean-field model, we consider the attosecond charge dynamics of electrons induced by strong-field laser pulses in two-dimensional Mott insulators. 
The necessity to go beyond single-particle approaches in these strongly correlated systems has made the simulation of two-dimensional extended materials challenging, and we contrast their resulting high-harmonic emission with more widely studied one-dimensional analogues. As well as considering the photo-induced breakdown of the Mott insulating state and magnetic order, we also resolve the time and ultra-high frequency domains of emission, which are used to characterize both the photo-transition, and the sub-cycle structure of the electron dynamics. This extends simulation capabilities and understanding of the photo-melting of these Mott insulators in two-dimensions, at the frontier of attosecond non-equilibrium science of correlated materials.
\end{abstract}
\maketitle

\section*{INTRODUCTION}
The non-linear response of a material under a strong driving field is increasingly being exploited for the high-harmonic generation (HHG) of coherent light, providing a powerful window into science on the shortest timescales \cite{RevModPhys.81.163}. This has allowed for recent advances, including probes of attosecond electron dynamics \citep{atto,atto2}, optical imaging of band structures \citep{PhysRevLett.115.193603}, and control and manipulation of quantum phases \citep{VO2,V02_2,floquet2,superconducting}, as well as answering some of the fundamental questions regarding the nature of decoherence in quantum systems \cite{PhysRevA.92.040502}. Simple phenomenological models for the charge dynamics which lead to this HHG have been proposed for atomic systems (the `three-step' model of tunnel-ionization, acceleration and recombination) \citep{atomic1,atomic2,atomic3,PhysRevLett.78.638}, with this more recently being extended to simple semiconductors, based on a fixed bandstructure and the build-up of inter and intra-band electronic currents \citep{solids1,solids2,solids4}, where HHG was demonstrated experimentally only in the last decade \citep{solids1,solids5,solids6}. However, the interplay of these temporal non-linear strong-field effects with non-perturbative strong electronic correlations, fundamentally changes the optical response of a material and enters a rich emergent regime of collective phenomena. This offers the possibility of emission of even higher harmonics due to coupling with these many-body interactions, and the proposal of engineering of photo-induced correlated phases \citep{floquet1,modificationHHG,mimicry1,mimicry2}. However, HHG and driving fields in strongly correlated materials is far from understood, and is emerging as a key research challenge in computational non-equilibrium science.

The introduction of strong correlation effects blurs the notion of well-defined band energies in the material, and complicates the simple interpretability of fixed bandstructure-based models. Regardless, the `single-particle' language of renormalized excitations to rationalize these phenomena remains important. This understanding is compounded by the difficulty in numerical simulation for these systems. In going beyond effective single-electron models \citep{modificationHHG}, reliable numerical investigations to date have been restricted to the consideration of one-dimensional (1D) systems \citep{spinHHG, thepaper, correlatedHHG, kondoHHG, 1DmottHHG}, where integrable models and matrix product states (MPS) are efficient, as well as dynamical mean-field theory approaches that work in an effective infinite-dimensional framework \citep{nonequilibriumHHG, mottHHG}. Exact diagonalization (ED) has also provided important insight in 1D \citep{thepaper}, but restrictions to small system sizes have stymied progress in investigating bulk limits. 
These two limits of dimensionality almost exclusively dominate the theoretical understanding of non-linear dynamics of correlated Mott insulators. Notwithstanding the important insight from these studies, many strongly-correlated systems of interest concern 2D layered materials, such as the cuprates or organic charge-transfer salts, which can give rise to a multitude of emergent correlated phases \citep{organicsaltsupercond,PhysRevB.99.184510}.

In this work we develop a time-dependent Monte Carlo method to extend numerical tools to understand the nature of HHG in strongly correlated 2D materials in large system limits. We consider how the phenomenology of these materials can vary from their 1D or infinite-dimensional analogues, and combine this with insight from ED to explore the effects of dimensionality and correlation on the time-averaged and time-resolved emission. From our simulations, we also map different interaction strengths to a effective single-particle model, exploring the extent to which single-particle modifications to the bandstructure in response to the local correlations (e.g. opening of Mott gaps) are responsible for the observed features of HHG, as well as the limitation of these one-body models where explicit many-body interactions are ignored. This aims to extend the understanding of HHG in these domains, and provide a tool for their numerical treatment in more realistic systems.
\begin{figure}
\centering
\includegraphics[width=1\linewidth]{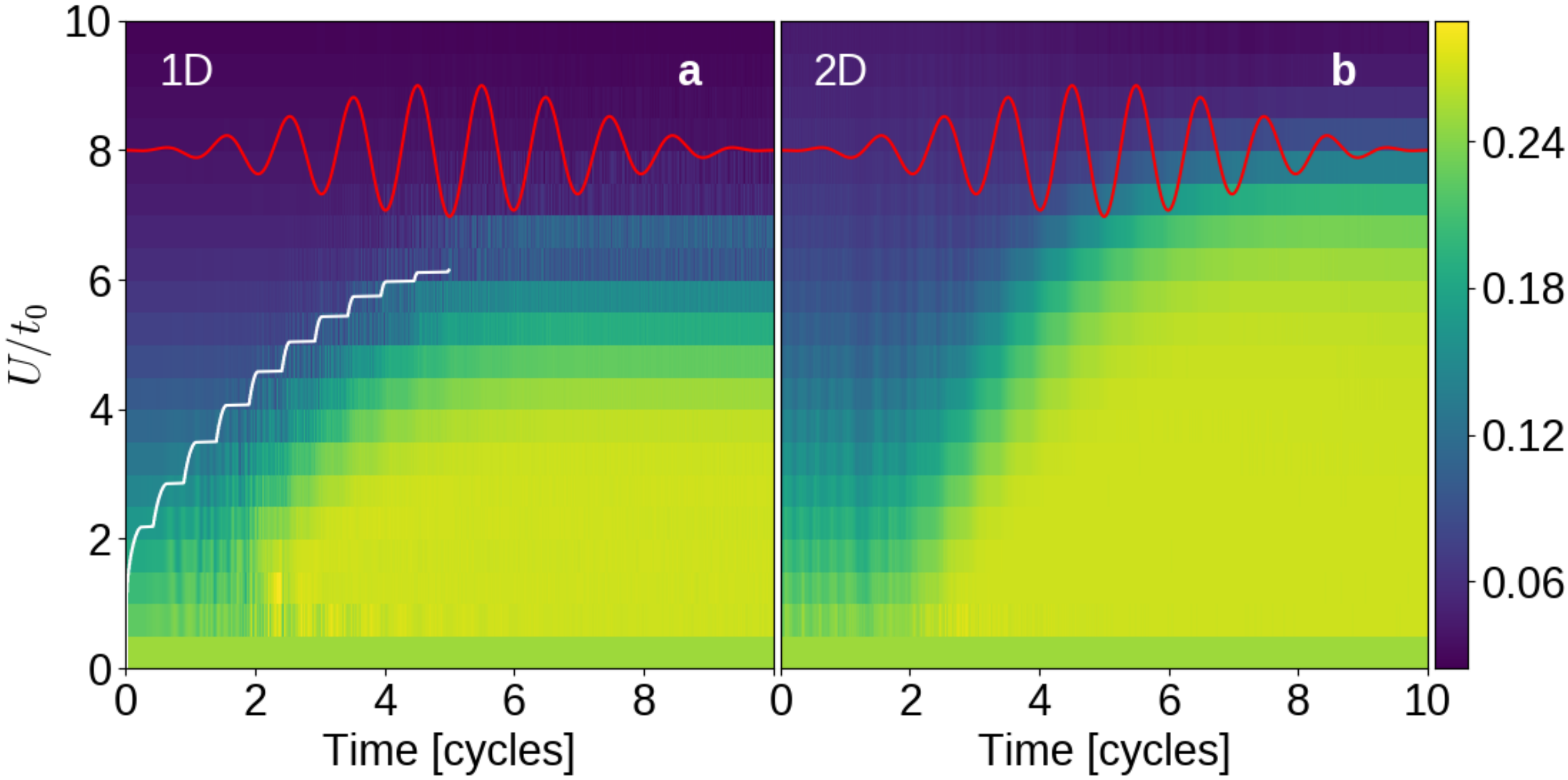}
\caption{\small{\textbf{Demonstration of photo-induced Mott transition.} Plot shows the time-evolution of the doublon density in the 1D \textbf{(a)} and 2D \textbf{(b)} Hubbard models under the driving pulse with temporal profile given in red, as $U/t_0$ is varied. Both systems are comprised of 12 site lattices, calculated by the exact propagation of the wave packet in time. The white line shows the time that the threshold field $E_{\text{th}}$ is exceeded, using analytic expressions for the Mott gap and coherence length in the thermodynamic limit, available in 1D from the Bethe ansatz.}}
\label{fig:observables}
\end{figure}
\section*{RESULTS AND DISCUSSION}
\subsection*{Photo-induced Mott transition}
As a paradigmatic framework for correlated Mott insulating systems, we work with the single-band Hubbard model, defined in the Methods section \citep{hubbard}. The system is subject to a monochromatic driving field with frequency $\omega_\text{L}$ and vector potential given by $A(t)=E_0/\omega_\text{L} \sin^2(\omega_\text{L} t/2N_\text{c}) \sin(\omega_\text{L} t)$, where $E_0$ is the peak amplitude and $N_\text{c}=10$ is the number of cycles. The electric field $E(t)=-\frac{dA(t)}{dt}$ is related to the Peierls phase by $\boldsymbol{\Phi}(t)=eaA(t) \mathbf{\hat{e}}$, where $a$ is the lattice constant, and $\mathbf{\hat{e}}$ controls the polarization of the pulse, which we take to be applied along the diagonal, $\mathbf{\hat{e}}=(1, 1)$. 
\toadd{We can connect the model to experimentally realizable THz pulse setups}, such that when the non-interacting low-energy bandwidth of the material is $\sim 2$~eV and lattice constant $\sim 4$\si{\angstrom}, the pulse considered has $E_0=10$ MV$\text{cm}^{-1}$ and $\omega_\text{L}=32.9$ THz, a physically realistic driving field with current technology in the mid-IR range \cite{Shalaby15}. \toadd{The specific model parameters that this corresponds to can be found in the Methods section, with these pulse and lattice parameters used for all simulations in this work.}

We first compare the effect of this driving on the 
average number of doublon-hole pairs \toremove{(DHP)}, defined as $D = \frac{1}{L} \sum_{i=1}^L \left \langle \hat{c}^\dagger_{i \uparrow} \hat{c}_{i\uparrow} \hat{c}^\dagger_{i \downarrow} \hat{c}_{i\downarrow} \right \rangle$, where $L$ is the total number of sites in the lattice. The Mott ground state of this model at half-filling has growing antiferromagnetic (AFM) order as interactions increase, and the concept of `melting' this N{\'e}el state via the elementary excitation of charge-carrying, non-magnetic \toremove{DHP} \toadd{doublon-hole pairs} is well established both experimentally and theoretically \citep{oka,oka2,VO2,V02_2,mott1}, and evidenced in \nref{fig:observables}. We consider the effect of dimensionality on this process, which requires excitations with energies above the Mott gap, $\Delta(U)$. For most correlation strengths considered here, this gap is considerably larger than the photon energy (in 1D, $\Delta >\omega_\text{L}$ for approximately $U > 2t_0$), and the melting therefore requires a highly non-linear excitation mechanism. The process responsible is determined by the Keldysh adiabaticity parameter, $\gamma=\hbar \omega_\text{L}/ \xi E_0$ \citep{oka,keldysh}, where $\xi(U)$ represents the correlation length of the excitations. Multi-photon absorption dominants for $\gamma \gg 1$, which is replaced by quantum tunneling when $\gamma \ll 1$, as in this region the Hubbard band energies are distorted by the static component of the field to such an extent that electrons can tunnel through the Mott gap to produce doublons. The 1D systems in this study are firmly in the latter regime, where DC dielectric breakdown is observed with threshold behaviour that is controlled by the peak field strength $E_0$, as can be seen in \nref{fig:observables}, and is independent of $\omega_\text{L}$. The threshold field strength required for breakdown is given by the Schwinger limit, $E_{\text{th}}=\Delta/2 e \xi$ \citep{oka}. Analytic expressions for the Mott gap and correlation length in the thermodynamic limit from the Bethe ansatz \citep{analytic,essler} allow for accurate predictions of $E_{\text{th}}$, which agree well with numerical results as seen in \nref{fig:observables} \citep{oka,thepaper}. In 1D, it is predicted that the threshold field is never exceeded ($E_{\text{th}}>E_0$) for $U > 6t_0$, above which the correlation is too large for this dynamical Mott transition to occur, as shown in \nref{fig:observables}.

A similar analysis of this photo-induced melting in 2D requires careful consideration to ensure that the restricted lattice sizes are still representative of the thermodynamic limit, and do not introduce spurious frustration caused by the boundary conditions (see supplementary materials). However, the results of \nref{fig:observables} are free from any other significant approximation. It is clear that the magnetic order is similarly melted, but now the transition continues to melt at higher correlation strengths than in 1D, as well as the spin correlation order demonstrating an earlier onset of melting in the pulse profile (see supplementary materials). This indicates a lower threshold field strength, arising from the the differences in bandstructure and correlation lengths between the models.
However, finite-size effects must be considered carefully, especially at small interactions where $\xi$ is large. To further investigate the effects of dimensionality, we develop approximate numerical approaches which can scale to larger system sizes and are more appropriate to probe the time and frequency dependence of the high harmonic emission in the thermodynamic limit.

\begin{figure*}
\includegraphics[width=1.0\linewidth]{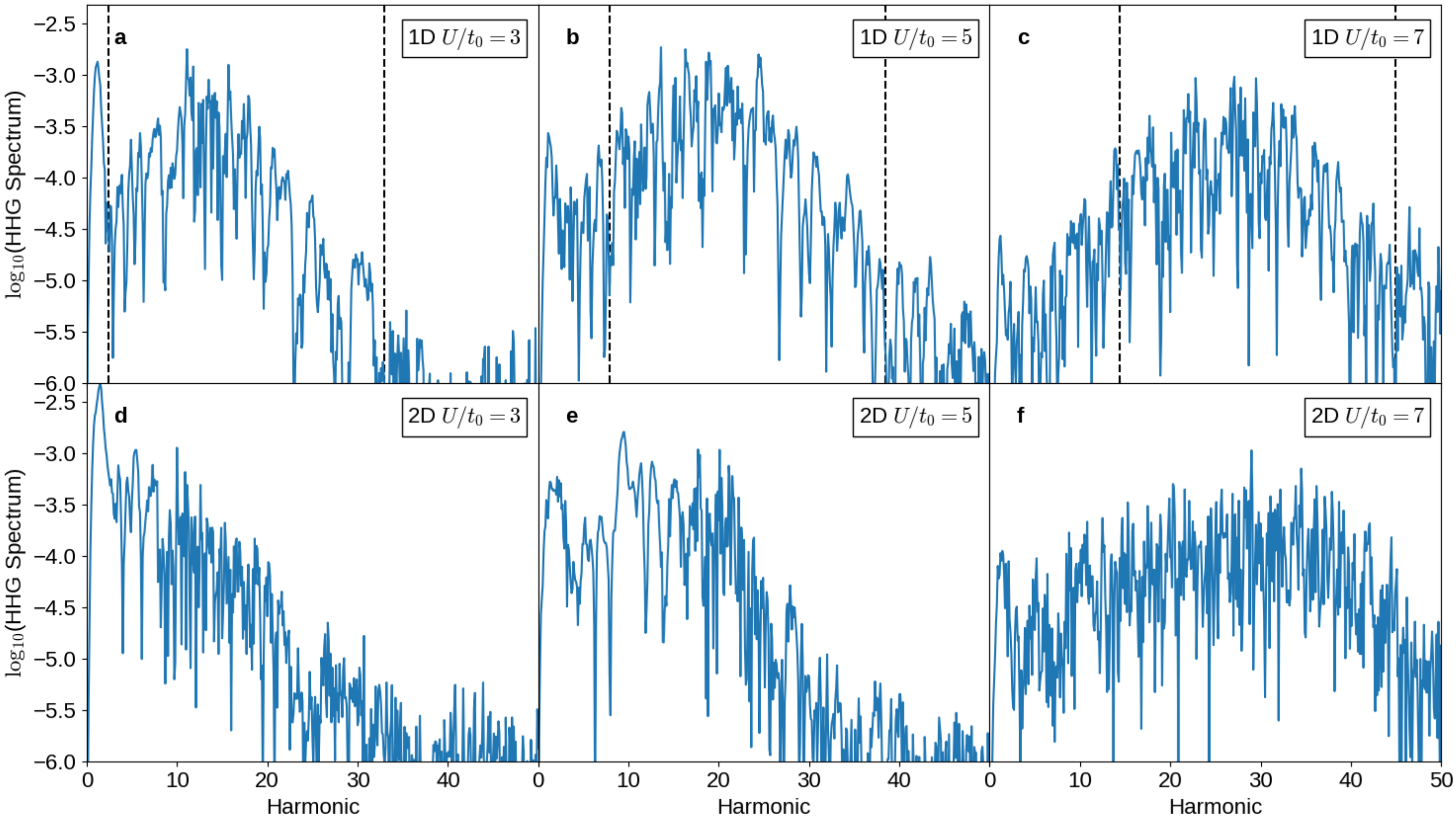}
\caption{\small{\textbf{Time-averaged spectrum against emitted high harmonic order.} HHG in 1D \textbf{(a-c)} and 2D \textbf{(d-f)}, for interaction strengths $U/t_0=3, 5, 7$, calculated by tVMC. The dashed lines in the 1D plots show the energy range $\Delta$ to $\Delta+8t_0$.}}
\label{fig:spectra}
\end{figure*}

For this, we have implemented the time-dependent variational Monte Carlo (tVMC) method \citep{tvmc}, which is combined with a projected Slater-Jastrow wave function form and stochastic reconfiguration to optimize the initial ground state within the {\tt mVMC} package \citep{mvmc,combined,SR1,SR2}. These wave functions have been shown to accurately capture a multitude of correlated phases of the Hubbard model, as well as the dynamics of quenches and other time-dependent phenomena \citep{tvmc,2DkondoHHG,SC,glassy,lightcone,quasilocality,neural1,neural3,neural4,ML1}. Independent samples in a 4th-order Runge-Kutta integrator were required for stable and accurate simulation of the HHG in these strong and rapidly changing fields, as discussed in the Methods section. \toadd{All tVMC and ED simulations were performed on $6 \times 6$ and 12 site lattices, respectively, with benchmarking of the accuracy of the tVMC method under these driving fields, as well as consideration of convergence with respect to lattice size for the HHG emission found in the supporting information.}

\subsection*{Time-averaged emission}
In \nref{fig:spectra}, we report the time-averaged HHG spectra, \toadd{defined by \cref{eq:Sw}}, resulting from both 1D and 2D simulations under the driving pulse, calculated using tVMC extended to larger system sizes. Results in 1D corroborate those previously found in the literature for ED of small lattice sizes \cite{thepaper}. These show low, odd-harmonic emission in the low interaction strength regimes (shown in supplementary materials), while stronger interactions rapidly give way to a broad spectrum of emission where peak harmonics increase linearly with interaction strength. These forms are rationalized in terms of distinct physical processes. The first relies on a fixed bandstructure picture, where the emission of sharp, distinct low-harmonics is a result of an intra-band current driven by dynamical Bloch oscillations, as is the dominant paradigm in semiconductor studies \citep{solids1}. \toremove{However, as interactions increase, in 1D this process is rapidly suppressed, with a significant Mott gap ($\Delta$) arising in the bandstructure to change the mechanism of emission.} \toadd{However, as interactions increase in 1D, a significant Mott gap ($\Delta$) arises in the bandstructure and changes the mechanism of emission.} Inter-band processes in this interacting regime dominate the emission, resulting from the driving field building polarization between the bands and forcing the creation of doublon-hole pairs and their subsequent recombination. For $U/t_0 \gg 1$, this gives rise to a broad band of high-harmonic emission, with the majority of the emission falling between $\Delta$ and $\Delta+8t_0$, which covers the range of \toremove{single momentum-conserving doublon-hole} \toadd{charge-conserving optical} excitations from the correlated ground state according to the Bethe ansatz \citep{thepaper,oka}. This forms a shifting interval of just over 30 harmonics, which peaks in the region of $N\sim U/\omega_\text{L}$. \toadd{This is accompanied by the suppression of lower harmonics, resulting from the upper Hubbard band becoming saturated by doublon-hole excitations, limiting the availability of quasi-momenta states for intra-band currents to form and thereby decreasing emission below the gap.}

In the non-interacting (metallic) limit, the emission in 2D is entirely proportional to that of 1D, and similarly driven by the intra-band current. However, as interactions increase, the presence of strong lower-harmonic emission persists, in contrast to 1D. \toremove{and despite the formation of a significant Mott gap.} \toadd{Specifically, there is emission surrounding the first harmonic at all correlation strengths approximately an order of magnitude more intense than the equivalent 1D model, despite the formation of significant Mott gaps.} This phenomenon is necessarily driven by a retention of significant levels of intra-band current, indicating a persistence of mobile collective charge carriers which avoid recombination to the AFM state over substantial timescales. This also agrees with \nref{fig:observables}, which shows the melting of AFM order even at large interaction strengths beyond the threshold field strength in 1D, facilitated by these persistent charge carrying quasiparticles which are absent at large interactions in 1D. We hypothesise that this is due to the ability of excitations to avoid each other in this higher dimensionality, which are therefore able to continue to undergo driven dynamical oscillations without recombination. \toremove{contributing to the broadening of the spectrum at large $U/t_0$. As well as this, we find the peak emission at intermediate interaction strength to be shifted to lower orders than their 1D counterparts.} \toadd{From a bandstructure perspective, this can be rationalized by noting that the single-particle bandwidth of the 2D model is twice that of 1D (representative bandstructures of these models can be found in the supporting information). This increased number of high-energy quasi-momenta states available in 2D ensures that the upper band does not become as saturated and can continue to support oscillations of itinerant doublons, leading to the persistence of low harmonic emission.}

\toadd{As $U/t_0$ increases in 2D, a separation develops between these intra-band current-induced lower harmonics, and higher frequency emission caused by localized doublon-holon recombination that extend up to tens of harmonics. At $U=5t_0$, interactions are sufficiently strong for local moments to begin to generate on the lattice, with the HHG shifted to lower orders than its 1D counterpart and dominated by a sharp peak near the 10th harmonic. The specifics of the van Hove singularities in the density of states for the 2D model, which are found closer to band edges in 2D than 1D (see supplementary materials), can explain this shift of peak intensity to lower harmonics, with a greater number of low-energy states around the Mott gap accessible in 2D.
This changes considerably in the strongly-correlated regime, at shown for $U=7t_0$. Here the spectrum broadens over more than 40 harmonics, caused by the large bandwidth in 2D which allows for a larger number of accessible high-energy states, combined with the continued persistence of intra-band currents at lower harmonics.}

\subsection*{Time-resolved emission}
\begin{figure}
\includegraphics[width=1\linewidth]{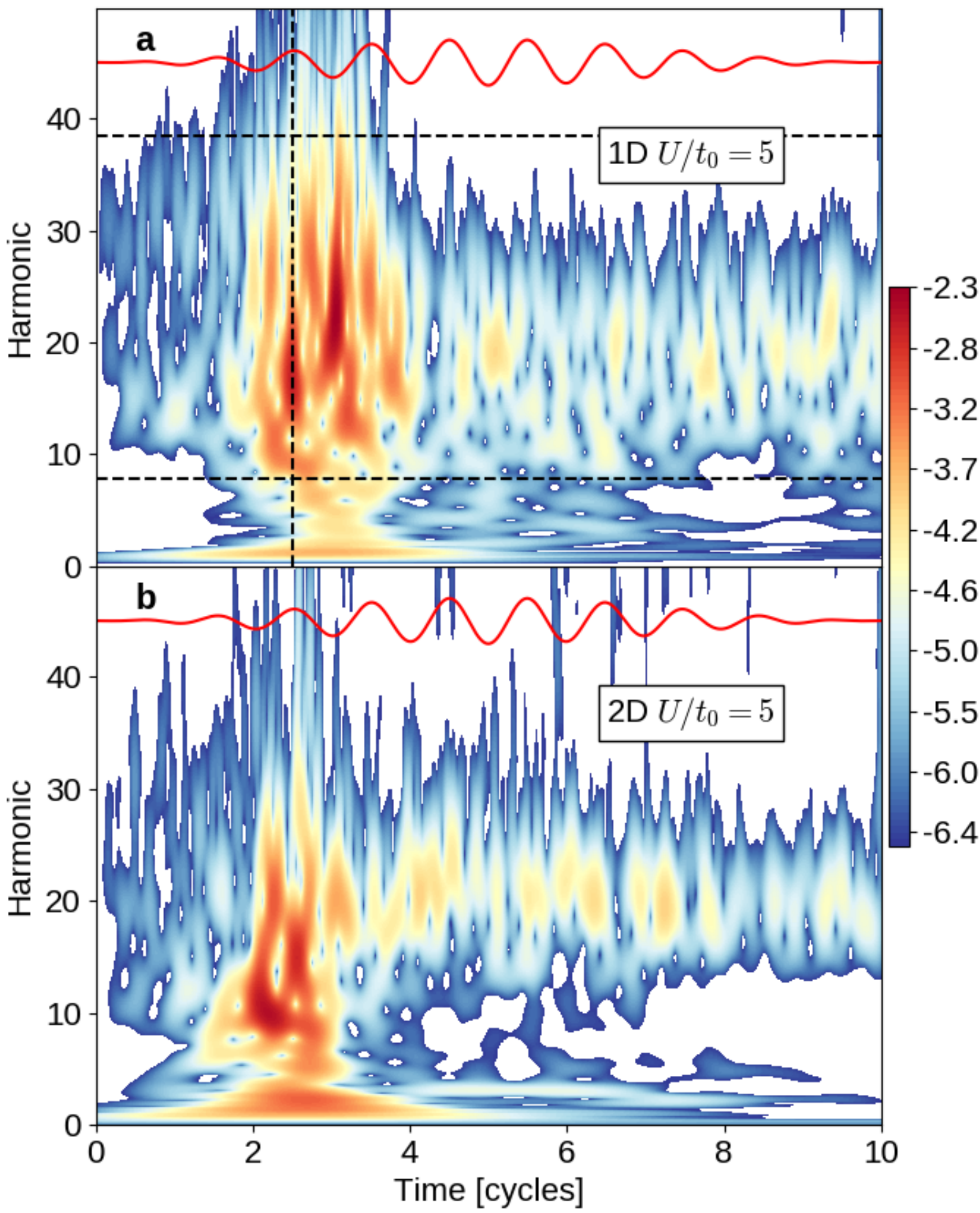}
\caption{\small{\textbf{Time-resolved high-harmonic emission.} Spectrograms in 1D \textbf{(a)} and 2D \textbf{(b)} at $U/t_0=5$, via tVMC. Colourscale denotes the log of the spectral emission intensity. The horizontal lines show the excitations $\Delta$ and $\Delta+8t_0$, and the vertical line shows the time when $E_{\text{th}}$ is reached in 1D.}}
\label{fig:spectro}
\end{figure}

The dynamics of the Mott transition and harmonic emission can also be time-frequency resolved on electronic timescales, \toadd{using the wavelet analysis defined by \cref{eq:Swt}} to extract the temporal emission amplitude. \Cref{fig:spectro} shows that the signature of the transition is a brief and intense period of emission, which peaks following the threshold $E_{\text{th}}$ being reached. This breakdown occurs within 2 field cycles in 1D and even more rapidly in 2D, and is followed by sub-cycle flaring at much lower intensities after the Mott gap has melted. In 2D, the peak emission occurs earlier, implying that the transitions happen more rapidly following irradiation, which is further corroborated by additional observations in supplementary materials, where the explicit melting of magnetic order is observed. We also find a stronger band of low-harmonic emission in 2D, corresponding to intra-band dynamical oscillations in the current. This emission starts almost instantly upon irradiation and persists for the first 4 cycles, providing almost all of the initial emission processes before significant recombination events can occur to induce higher harmonics.  
The threshold field is proportional to the Mott gap, which makes $\Delta(U)$ an important factor in whether the transition can occur and how quickly after irradiation. \toadd{That the gap for a given interaction strength is smaller in 2D can explain many of these observed differences, in particular the lower threshold field strength observed in \nref{fig:observables} giving rise to a melting of Mott order at higher values of the interaction, as well as the earlier onset of transitions observed in \nref{fig:spectro}. The Mott gaps in these different dimensionalities can be seen in the supplementary materials, and are in qualitative agreement with cluster perturbation theory results \cite{gap, gap_2}.} 

\begin{figure}
\includegraphics[width=1.0\linewidth]{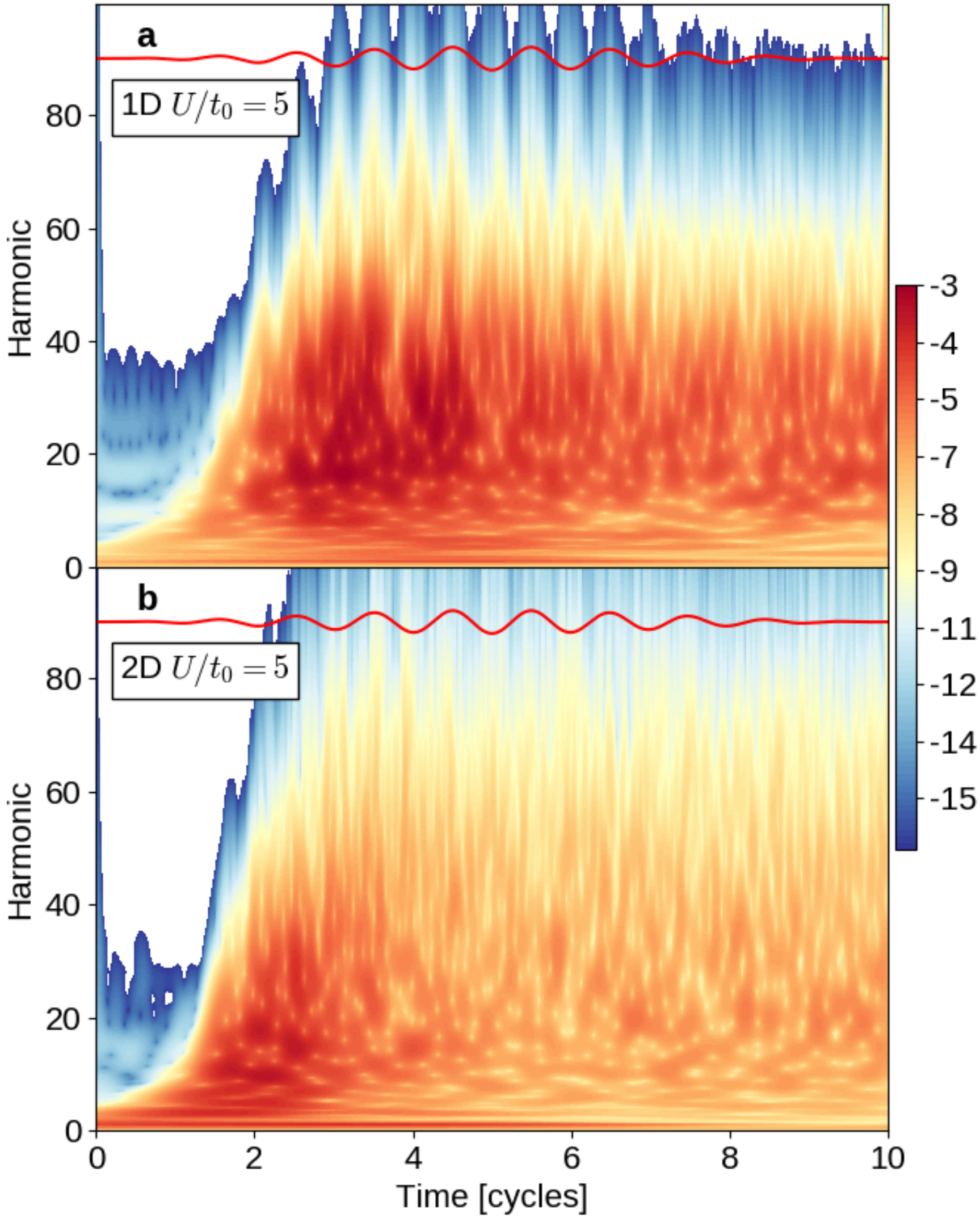}
\caption{\small{\textbf{Time-resolved high-harmonic emission.} Spectrograms showing the log of the emission in 1D \textbf{(a)} and 2D \textbf{(b)}, at $U/t_0=5$. \toadd{The scales are chosen to highlight the high-frequency, low-intensity emission required to observe sub-cycle characteristics not seen in the emission ranges of Fig.~\ref{fig:spectro}.}}}
\label{fig:exactspectro}
\end{figure}

The time-resolved spectrum also exhibits alignment between electric field peaks and the sub-cycle emission profiles, only \toadd{visible at intensities orders of magnitude smaller than \nref{fig:spectro}, where these features are masked by the high-intensity} emission derived from the dynamical Mott transition. \Cref{fig:exactspectro} \toadd{allows this structure to be seen at intensities less than $\sim 10^{-6}$ and emission frequencies significantly greater} than the scale of $\sim U/\omega_\text{L}$. Due to the low intensity of this emission, the tVMC method struggles to resolve these \toadd{sub-cycle} details against the inherent stochastics, and so we show exact results benchmarked against the larger systems accessible with tVMC to ensure no spurious finite size effects are introduced into our conclusions. In 1D there is highly regular emission, best understood in terms of an extended three-step model \citep{oka,1DmottHHG}. In this picture in 1D, \toremove{DHP} \toadd{doublon-hole} excitations are created through tunneling, accelerated away by the field and then returned as the field oscillates back. The resulting recombination causes two emission peaks per cycle which are aligned with crests in the field strength, and primarily occur during the pulse's central few cycles where $E(t)$ is largest. In 2D, the HHG has a more featured sub-cycle structure, including multiple flares per half cycle which do not only coincide with peak field, and which persist longer into the pulse profile. This complexity likely arises from the larger number of trajectories that these correlated excitations can take in this higher dimensionality, leading to a more complicated structure of allowable constructive and destructive interference patterns within each cycle. These specific dynamical trajectories can be elucidated via an analysis of the complex phase of the dipole acceleration, which we will consider further in future work. 

In both 1D and 2D, the observed characteristics of the HHG hold over a broad range of pulse and lattice parameters. In the frequency domain, properties such as cutoff and peak harmonic region do not vary significantly with $E_0$, which arises as all possible contributing excitations are already accessible \citep{oka,thepaper}, provided the field strength is sufficient to melt the Mott order via dielectric breakdown ($E_0 \ge E_{\text{th}}$). This holds in both dimensionality regimes, with the dominant mechanism of tunnelling via dielectric breakdown remaining valid. However, both $E_0$ and the effective lattice constant do affect how rapidly the threshold field is reached, thereby changing the onset of the dynamical phase transition. Due to the tunnelling mechanism, the field frequency does not considerably effect this transition \citep{oka}, but rather shifts the emission peaks to larger harmonics as the frequency is decreased.

\subsection*{Effective single-particle model}
The HHG of a Mott insulator is clearly determined by the correlated dynamics of the many-electron problem. However, discussion of the dominant mechanisms generally considers the renormalized one-body dynamics described by the ground state bandstructure, with e.g. the effects of the Mott gap a central quantity for correlated materials. We aim to investigate the extent to which a characterization of this single-particle bandstructure, as modified by the correlations, can account for some of the features we observe in the HHG. This will allow us to extract the simple bandstructure-driven features of the response, in isolation from explicit many-body correlated features which are harder to characterize. 

We map the driven system onto a simplified non-interacting model shown explicitly in the Methods section, introducing fictitious degrees of freedom which couple locally to each physical site of the lattice with a strength parameterized by the scalar-valued $V(U)$. These fictitious degrees of freedom hybridize with the physical lattice in such a way as to mimic the band splitting caused by the correlations of the interacting lattice. The parameter $V(U)$ is optimized for each interaction strength by matching the first spectral moment of the particle and hole density of states of the ground state, given by $\langle \Psi | \hat{c}^{(\dagger)}(\hat{H}-E)\hat{c}^{(\dagger)}|\Psi\rangle$, between the non-interacting model and the correlated system as described by variational Monte Carlo (VMC) \citep{auxillary,auxillary2,sriluckshmy2021fully}. After tracing out the fictitious degrees of freedom, this can introduce a second band and open an effective Mott gap in the bandstructure of the non-interacting model, $\Tilde{\Delta}(U)$, with the fictitious space representing a local self-energy, comprising a single pole with spectral weight $|V(U)|^2$ at the Fermi level. This therefore represents a simplified {\em dynamical} mean-field model, with the physical space dynamics remaining norm-conserving under the applied pulse. \toadd{This is similar to the physics captured by the `Hubbard I' approximation, albeit in that approach the effective self-energy pole is constructed from the atomic solution, rather than than the VMC ground state solution to the full correlated lattice in this instance \cite{hubbard1}. Similar-in-spirit `semi-conductor' models designed to mimic the correlated bandstructure in a single-particle model have also been considered elsewhere, in order to rationalize qualitative features of HHG emission \cite{mottHHG}. A description of the density of states and resulting band structures from this model can be found in the supplementary material.}

While Mott physics can indeed emerge from this model, \toadd{with the band gaps in qualitative agreement with results from cluster perturbation theory (2D) and the Bethe Ansatz (1D) \cite{gap, gap_2}, there are some significant limitations}. Firstly, the effective minimal self-energy is fixed by the ground state correlations, and does not change with time in response to the pulse, \toadd{as opposed to the explicitly non-equilibrium self-energy in the case of} e.g. non-equilibrium DMFT \cite{nonequilibriumHHG,mottHHG}. Secondly, these auxiliary sites are only locally coupled, precluding a momentum-dependent, long-range modification to the effective correlations. Finally, no spin-dependent terms are optimized, so the system remains in a paramagnetic phase with no effective spin-dependent Heisenberg interactions arising from consideration of an explicit Hubbard term. This in itself \toadd{precludes the formation of local moments}, and therefore the melting of this order which features so prominently in the results of the time-frequency resolved emission of \nref{fig:spectro}.

\begin{figure}
\includegraphics[width=1.\linewidth]{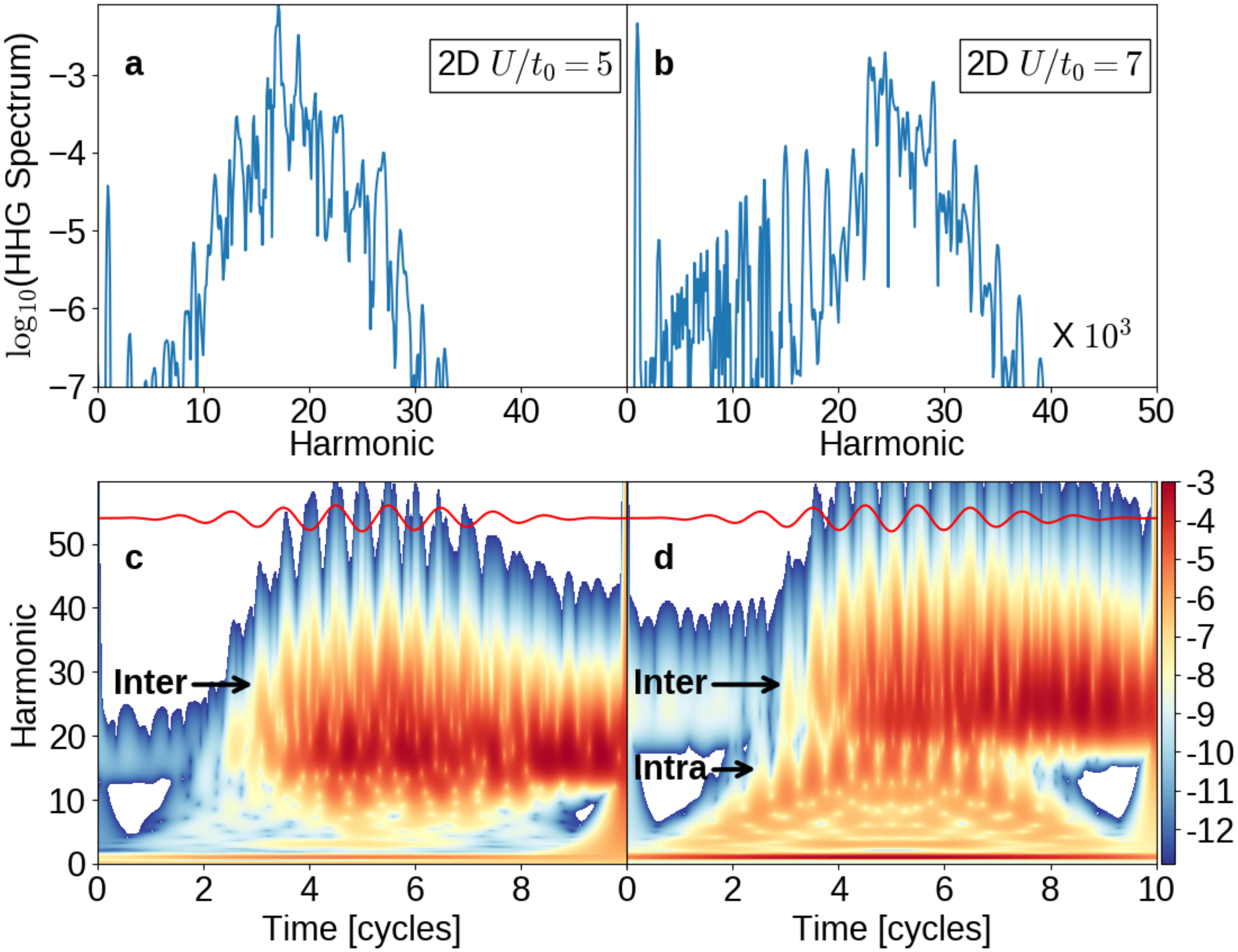}
\caption{\small{\textbf{HHG in the effective single-particle model.} Time-averaged and time-resolved HHG in 2D, describing interaction strengths $U/t_0=5$ \textbf{(a,c)} and $U/t_0=7$ \textbf{(b,d)} (multiplied by $10^3$), corresponding to optimized values of $V=2.1$ and $V=3.1$, respectively. The $U/t_0=7$ spectrogram shows a splitting of the emission into two sectors according to its origin in intra- or inter-band processes, with these emission bands labelled on the plot.}}
\label{fig:2D_nonint}
\end{figure}

Nevertheless, the single-parameter simplicity of this bandstructure-engineered model is appealing for its interpretability, as well as being simple to solve in the true bulk system limit. It manages to capture some of the broad features of the spectra arising from intra- and inter-band processes, which can now be directly attributed to renormalized single-particle dynamics on this modified bandstructure. At $U/t_0=0$ the model is exact ($V=0$). \Cref{fig:2D_nonint} shows that as the effective interactions increase, the spectrum becomes dominated by a broad band of emission resulting from recombination events following inter-band excitations. This emission is mostly confined to harmonics between $\Tilde{\Delta}$ and the largest excitation, and the peak emission amplitude goes as $\sim U/\omega_\text{L}$. These are properties shared by the explicitly correlated spectra in \nref{fig:spectra}, particularly in 1D, and is a clear demonstration of the validity of simple bandstructure and Mott gap arguments for the main time-averaged emission profiles. Furthermore, the model becomes increasingly accurate as $U/t_0\rightarrow \infty$, where the system is further from dielectric breakdown. In this limit the upper band becomes sparsely populated, allowing the formation of intra-band currents which cause the spectrogram to develop a clear separation between inter- and intra-band driven emission, as seen in \nref{fig:2D_nonint} for $U/t_0=7$, both of which cause field-synchronized HHG, with the former at high harmonics and the latter at lower ones. This also occurs in 1D and 2D systems at sufficiently large interactions, and numerical demonstration of this equivalence to the fully correlated dynamics in this limit are given in supplementary materials. This shows that the model is more appropriate for use in the high-$U/t_0$ limit, where a single-particle description within a static bandstructure picture is sufficient to capture most features of the sub-cycle dynamics. It is clear therefore that bandstructure changes alone leave important imprints on the HHG in correlated materials.

However, both the absolute emission amplitude, as well as the relative amplitudes between the low-harmonic intra-band and high-harmonic inter-band emission are not faithfully reproduced, with an exponential decay of emission intensity for increasing effective interaction strengths due to the persistence of the Mott gap. Overall, the model reproduces some important features of the correlated spectra but unsurprisingly struggles to capture the complexity of the interfering trajectories in 2D due to its lack of explicitly correlated dynamics. At effective $U/t_0=5$, low harmonics are suppressed, which is observed in the 1D spectra but not in 2D where these low harmonics are retained. Furthermore, the lack of threshold behaviour introduces a sensitivity to the pulse parameters $E_0$ and $a$ which do not exist in the fully correlated model. Improvements can be made by matching both spin and charge gaps, as well as introducing a time-dependence into this effective self-energy of the auxiliary system. However, these will come at the cost of increased complexity and difficulty in interpretation of the model.

\section*{SUMMARY}
We tackle the challenging simulation of high harmonic generation in two-dimensional strongly-correlated Mott insulators. By application of a time-dependent stochastic technique, we are able to access the system sizes required to elucidate the emission profiles in both the temporal and frequency domains, as well as compare and contrast the emission with lower-dimensional models. We also map the system to a simplified, single-parameter dynamical mean-field picture, which allows for the salient Mott-gap driven emission to be considered in isolation. These tools and insight will further the study of non-equilibrium correlated electron systems on their natural timescales, and extend the theoretical understanding of high-harmonic probes of these critical many-body charge dynamics in attosecond science. 

\section*{METHODS}
\subsection*{Model}
We consider 1D and 2D (square lattice) Hubbard model with periodic boundary conditions at half-filling,
\begin{align} \label{eq:Cham}
\begin{split}
\hat{\mathcal{H}}(t)=&-t_0 \sum_{\langle ij \rangle \sigma}  \left \{ \text{e}^{-\text{i}\Phi(t)}\hat{c}^\dagger_{i\sigma} \hat{c}_{j\sigma} + \text{e}^{\text{i}\Phi(t)}\hat{c}^\dagger_{j\sigma} \hat{c}_{i\sigma} \right \} \\ & +  U \sum_i \hat{n}_{i\uparrow} \hat{n}_{i \downarrow} ,
\end{split}
\end{align}
\toadd{where $\langle ij \rangle$ denotes a hopping to nearest neighbours (such that in 1D these sites are arranged in a chain), and} variations in the effective screened Coulomb interaction ($U/t_0$) dictate the importance of the correlated physics, and which can act as a proxy for other external parameters to induce quantum phase transitions. Simple $sp$-band systems such as graphene map to $U\sim 1.5t_0$ \citep{PhysRevB.57.6884,DFTU}, while strongly correlated cuprate parent materials of high-temperature superconductivity are modelled with $U\sim 8t_0$ \citep{cuprate1,cuprate2}. \toadd{Taking $t_0=1$, the pulses considered in this work correspond to model parameters $\omega_\text{L}= 0.262$, $E_0=5.325$, and $a=1.444$.}

The electric current operator is defined as
\begin{equation} \label{eq:current}
\hat{\mathcal{J}}(t)=-\text{i} e a t_0 \sum_{\langle ij \rangle \sigma}  \left \{ \text{e}^{-\text{i} \Phi(t)}\hat{c}^\dagger_{i\sigma} \hat{c}_{j\sigma} - \text{e}^{\text{i}\Phi(t)}\hat{c}^\dagger_{j\sigma} \hat{c}_{i\sigma}\right \} ,
\end{equation}
where the emitted HHG is defined from the charge acceleration as 
\begin{equation} \label{eq:Sw}
    \mathcal{S}(\omega)= \left |\mathcal{FT} \left \{ \frac{dJ(t)}{dt} \right \} \right|^2,
\end{equation}
with $J(t)=\langle \hat{\mathcal{J}}(t) \rangle$. The emission can also be resolved in time, allowing the Mott transition and electron dynamics to be directly observed. This was done using wavelet analysis, which takes the convolution of a wavelet $\psi(\eta)$ with the acceleration $a(t_n)=dJ(t_n)/dt$ for $n=0,...,N-1$ \citep{wavelet}
\begin{equation} \label{eq:Swt}
    a(s, t_n) = \sum_{t_{n'}=0}^{N-1} a(t_{n'}) \psi^* \left [ \frac{(t_{n'}-t_n)\Delta t}{s} \right ],
\end{equation}
where $s$ are the wavelet scales and $\psi(\eta)$ was taken as a Morlet wavelet. 

\subsection*{Exact diagonalization}
For small systems the Schr{\"o}dinger equation 
\begin{equation}
  \frac{d\ket{\Psi( t)}}{dt} =-\text{i} \hat{\mathcal{H}}(t) \ket{\Psi( t)},
\end{equation}
can be solved numerically without approximations in the expressibility of the state at any time. The initial state $\ket{\Psi(0)}$ is taken to be the ground state, calculated with exact diagonalization (ED) \citep{pyscf}. The wavefunction $\ket{\Psi(t)}$ is then evolved forward in steps $\Delta t$ using the 4th-order Runge-Kutta method (RK4). This uses a series of recursive steps
\begin{align} \label{eq:rk4}
\ket{k_1} &= -\text{i} \Delta t \hat{\mathcal{H}}(t) \ket{\Psi(t)} \\
\ket{k_2} &=  -\text{i} \Delta t \hat{\mathcal{H}}(t+\Delta t/2) \left \{ \ket{\Psi(t)} + \ket{k_1}/2  \right \}\\
\ket{k_3} &=  -\text{i} \Delta t \hat{\mathcal{H}}(t+\Delta t/2) \{\ket{\Psi(t)} + \ket{k_2}/2  \}\\
\ket{k_4} &=  -\text{i} \Delta t \hat{\mathcal{H}}(t+\Delta t) \{\ket{\Psi(t)} + \ket{k_3}  \} \label{eq:rk42}
\end{align} 
with update 
\begin{equation}
\ket{\Psi(t+\Delta t)} =  \ket{\Psi(t)} + \frac{1}{6}\left (\ket{k_1}+2\ket{k_2}+2\ket{k_3}+\ket{k_4}\right).
\end{equation}
\toremove{All ED simulations in this work are restricted to 12 site systems.} \nref{fig:exactspectro} uses a square lattice, but \nref{fig:observables} uses a tilted lattice to better capture the doublon density in the thermodynamic limit (see supplementary materials). 

\subsection*{Trial wavefunction}
The dimension of the full Hilbert space increases exponentially with the number of sites in the lattice, which necessitates the use of alternative methods to probe the thermodynamic limit. Instead of working with the full many-body wavefunction, tVMC uses a trial wavefunction of the form $\ket{\Psi}=\ket{\Psi(\alpha_1,...,\alpha_p)}$, which depends on a set of $p$ complex-valued variational parameters. The wavefunction used in this work was \citep{mvmc,tvmc}
\begin{equation}
\ket{\Psi(t)}=\mathcal{L}_K \mathcal{L}_P  \mathcal{P}(t)\ket{\phi(t)},
\end{equation}
where $\mathcal{L}_K$ is the momentum number projector, $\mathcal{L}_P$ is the point-group symmetry projector, $\mathcal{P}(t)$ is a Jastrow-Gutzwiller correlation factor, and the Pfaffian wavefunction is given by $\ket{\phi(t)}=\left ( \sum_{i,j}^{L} f_{ij}(t) \hat{c}^\dagger_{i\uparrow} \hat{c}^\dagger_{i\downarrow} \right)^{N_\text{e}/2} \ket{0}$.

\subsection*{Stochastic methods}
The parameters for the Hubbard ground state were optimized simultaneously using the stochastic reconfiguration (SR) method \citep{SR1}, which was then evolved in real time using the time-dependent variational Monte Carlo (tVMC) method \citep{tvmc}. 

Both methods use the time-dependent variational principle \citep{TDVP1,TDVP2}. In real time, at each step the parameters are calculated to minimise \citep{tvmc}
\begin{equation} 
\underset{\boldsymbol{\upalpha}}{\text{min}} \Big | \Big | \text{i} \frac{d\ket{\Psi(\boldsymbol{\upalpha}(t))}}{dt}-\mathcal{H}\ket{\Psi(\boldsymbol{\upalpha}(t))} \Big | \Big |. 
\end{equation}
This leads to a system of equations for the trajectory of the parameters
\begin{equation} \label{eq:realsystem}
\frac{d\alpha_k}{dt}=-\text{i} \sum_m S^{-1}_{km} g_m,
\end{equation}
where Markov chain Monte Carlo is used to calculate the matrix \textbf{S} and vector \textbf{g}.

\Cref{eq:realsystem} was propagated forward using RK4, and at each timestep the expectation value $J(t)=\langle \hat{\mathcal{J}}(t) \rangle$ of the current operator was calculated using VMC. In SR, the ground state is found by evolving \cref{eq:realsystem} with the Euler method in imaginary time within the natural metric of the parameterization, under the substitution $\tau = \text{i}t$. \toremove{All tVMC simulations were done with 36 site lattices.} 

\subsection*{Effective single-particle model}
The Hamiltonian is given by
\begin{align} \label{eq:CBM}
\begin{split}
\hat{\mathcal{H}}(t)=&-t_0 \sum_{\langle ij \rangle \sigma}  \left \{ \text{e}^{-\text{i}\Phi(t)}\hat{c}^\dagger_{i\sigma} \hat{c}_{j\sigma} + \text{e}^{\text{i}\Phi(t)}\hat{c}^\dagger_{j\sigma} \hat{c}_{i\sigma} \right \} \\ &+  V(U) \sum_{i \sigma}  \left \{ \hat{c}^\dagger_{i\sigma}\hat{f}_{i\sigma} + \hat{f}^\dagger_{i\sigma} \hat{c}_{i\sigma} \right \}
\\ &-\frac{U}{2} \sum_{i\sigma} \left \{ \hat{c}^\dagger_{i\sigma}\hat{c}_{i\sigma} + \hat{f}^\dagger_{i\sigma}\hat{f}_{i\sigma} \right \}.
\end{split}
\end{align}
The second term couples each physical site to its respective auxiliary one via the Fermionic operators $\hat{f}^{(\dagger)}$. This replaces the interaction term of the Hubbard Hamiltonian, and introduces a gap via the single parameter $V(U)$. The advantage of this method is that the model is entirely non-interacting, so the system's dynamics can be exactly and rapidly calculated using its time-dependent one-body density matrix. This was done once again using the RK4 method. 

The mapping between interacting (at a given $U/t_0$) and non-interacting systems is achieved by choosing $V(U)$ to match the mean of the particle and hole distributions of the density of states. This is first calculated for the correlated system using VMC, by writing it as an expectation of the ground state wave function \citep{auxillary,auxillary2,sriluckshmy2021fully}
\begin{align}
T =& \ \frac{1}{L} \text{Tr} \big [ \langle \Psi_0 | \hat{c}^\dagger_{\beta} [\hat{c}_{\alpha}, \hat{\mathcal{H}}] | \Psi_0  \rangle \big ] \\ \label{eq:T}
=& -\frac{t_0}{L} \sum_{\langle \alpha, \beta \rangle} \mathcal{D}_{\alpha \beta} + U\left( 2D - \frac{1}{2} \right)
\end{align}
where $\mathcal{D}_{\alpha \beta}$ is the one-body density matrix summed over nearest neighbour sites $\alpha$ and $\beta$, while $D$ is the ground state double occupancy on each site. The parameter $V$ is then optimized so that $T$ in the physical system (once the auxiliary system is traced out) matches that of the correlated system. At a given $V$, the mean-field moment is given by
\begin{equation} \label{eq:Tmf}
T = \frac{2}{L} \sum_\alpha \sum_{\epsilon_j<\mu} C_{\alpha j} C_{\alpha j} (\epsilon_j - \mu),
\end{equation}
where $\mu$ is the chemical potential and $(\mathbf{C};\epsilon)$ are the eigenvectors and eigenvalues of the fictitious composite system's Hamiltonian, and $\alpha$ is restricted to indices of the physical (not auxiliary) system. The auxiliary system therefore acts as a entanglement bath to change the one-body dynamical properties of the model.

\section*{DATA AVAILABILITY}
Data available on request from the
authors.

\toadd{\section*{CODE AVAILABILITY}
The codes used to perform these calculations are available on request from the authors.}


\section*{ACKNOWLEDGEMENTS}
The authors thank Gerard McCaul for helpful discussions on the manuscript.
G.H.B. gratefully acknowledges support from the Royal Society via a University Research Fellowship, as well as funding from the European Research Council (ERC) under the European Union's Horizon 2020 research and innovation programme (grant agreement No. 759063). We are also grateful to the UK Materials and Molecular Modelling Hub for computational resources, which is partially funded by EPSRC (EP/P020194/1 and EP/T022213/1), as well as A.Z. receiving funding from the Royal Society (RGS/R1/211053)
C.O. acknowledges funding by the Engineering and Physical Sciences Research Council (EPSRC) through the Centre for Doctoral Training ``Cross Disciplinary Approaches to Non-Equilibrium Systems" (CANES, Grant No. EP/L015854/1).

\section*{AUTHOR CONTRIBUTIONS}
G.H.B. devised and supervised the project. C.O. performed all simulations, which were analyzed by all authors. G.H.B. and C.O. wrote the paper, with comments and contributions from all authors.

\section*{COMPETING INTERESTS}
The authors declare no competing interests.

\section*{ADDITIONAL INFORMATION}

{\bf Supplementary information} The online version contains supplementary material
available at [link to be added].

\noindent{\bf Correspondence} and requests for materials should be addressed to G.H.B.

\end{document}